\documentclass[11pt,a4, reqno]{article}

\usepackage{amssymb}
\usepackage{amsmath}

\begin{document}

\title{Spin Glasses: a Perspective}

\author{David Sherrington\\Rudolf Peierls Centre for Theoretical Physics, 
\\University of Oxford, 1 Keble Rd., Oxford OX1 3NP, UK}

\maketitle

\begin{abstract}
A brief personal perspective is given of issues, questions, formulations, methods, some answers 
and selected extensions posed by the spin glass problem, showing how considerations of an 
apparently insignificant and practically unimportant group of metallic alloys stimulated an 
explosion of new insights and opportunities in the general area of complex many-body systems 
and still is doing so.
\end{abstract}

\section{Introduction}
What are spin glasses? The answer to this apparently innocuous question has evolved from 
an initially obscure, if interesting, small special class of metallic alloys to ones concerned 
with the globally pervasive issue of the understanding of emergent complex behaviour in 
many-body systems, the development of new mathematical, simulational and conceptual tools, 
new experimental protocols, new algorithms and even a new class of mathematical probability 
problems. In this article I shall review some of this history and try to expose some of the 
key issues, challenges, solutions and opportunities of the topic.

\section{Random magnetic alloys}
The story starts with magnetic aspects of metallic alloys. In the early 1960's there was 
much interest in the solid state physics community in the behaviour of isolated impurities 
in metals, first with the formation of local magnetic moments on magnetic metal impurities 
in non-magnetic hosts (Anderson 1961) and then with the strong coupling of a localized 
moment to the conduction electrons at low temperatures and its consequences for the 
electrical resistivity (Kondo 1964). The later 
60's and early 70's saw the emergence of interest in the effects of inter-impurity
correlations through spin glasses (see e.g.\ Coles 1983 and M\'ezard et al 1987) and the
Kondo lattice (Doniach 1977).

The appellation \lq\lq spin glass\rq\rq\ is due to Bryan Coles in the late 60's to label 
the low temperature state of a class of substitutional magnetic alloys, typified by
\textbf{Cu}Mn or \textbf{Au}Fe, with finite concentrations of the magnetic ions Mn or Fe
in the non-magnetic hosts Cu and Au. The reason for the name is two-fold, first that in
the state the magnetic moments (traditionally called \lq\lq spins\rq\rq) on the magnetic
ions seem to freeze in orientation but without any periodic ordering (so conceptually
reminiscent of the amorphous freezing of the locations of atoms in a conventional
(structural) glass), and secondly that the low temperature specific heat is linear in $T$,
again a feature of conventional glasses. Experiments at that time indicated a non-sharp
onset of the state as the temperature was reduced from the paramagnetic one, suggesting a
rapid onset of sluggishness but not a phase transition, again as believed to be
characteristic of glasses.  There were attempts to understand the behaviour in the 60's
but nothing very extensive.

But then more accurate experiments in the early 1970's exposed a new source of theoretical 
interest, an apparently sharp phase transition signalled by a cusp in the magnetic 
susceptibility when external magnetic fields were kept very small  (Cannella and Mydosh 
1972). This had to be a new type of phase transition and therefore worthy of extra notice. 
But still theoretical work was minimal until Edwards and Anderson (1975) produced a paper 
that at one fell swoop recognised the importance of the combination of frustration and 
quenched disorder as fundamental ingredients, introduced a more convenient model, a new 
and novel method of analysis, new types of order parameters, a new mean field theory, 
new approximation techniques and the prediction of a new type of phase transition apparently 
explaining the observed susceptibility cusp. This paper was a watershed.

Edwards and Anderson's new approach was beautifully minimal, 
fascinating and attractive but also their analysis was highly novel and sophisticated,
involving radically new concepts and methods but also unusual and unproven ans\"atze, as
well as several different approaches. So it seemed sensible to look for an exactly
soluble model for which their techniques could be verified. Such a model was suggested by
Sherrington and Kirkpatrick (1975). It extends the Edwards-Anderson model, in which
exchange interactions are range-dependent and effectively short-range, to one with
interactions between all spins, chosen randomly and independently from an intensive
distribution (and so `infinite-ranged' but not uniform). It offered the
possibility of exact solution in the thermodynamic limit and an exact mean field theory,
in analogy but subtle extension of the infinite-range ferromagnet  for which 
na\"\i ve
mean field theory is correct. Study of the SK model, or the mean field theory of the EA
model that it defines, has proven highly non-trivial and instructive, and opened many new
conceptual doors. It has also proven to be an entry point to many applications and
extensions, which are still ongoing.

\subsection{More details}

\subsubsection{Experimental spin glasses}
Let me be more explicit. The original experimental spin glasses can be characterised by Hamiltonians
\begin{equation}
\label{eq1}
H=-\frac12\sum_{i,j}J(\mathbf R_i-\mathbf R_j)\mathbf S_i\cdot\mathbf{S}_j ,
\end{equation}
where the $i$, $j$ label magnetic ions with Heisenberg spins $\mathbf S_i$ and locations 
$\mathbf R_i$ and $J(\mathbf R)$ is an exchange interaction which oscillates in sign as a 
function of the spin separation. In metallic systems the origin of $J(\mathbf R)$  is the 
Ruderman-Kittel-Kasuya-Yoshida (RKKY) interaction. In the original alloys the disorder 
is substitutional on a lattice.

\subsubsection{Edwards-Anderson}
What Edwards and Anderson (correctly) surmised was that the important aspect of \eqref{eq1} 
is the combination of frustration, corresponding to the fact that the spins receive 
conflicting relative ordering instructions (as a consequence of the oscillation of the 
exchange with separation), 
and the quenched disorder in the location of the spins. For theoretical convenience they 
effectively replaced the Hamiltonian by one that can be written as
\begin{equation}\label{eq2}
H=-\frac12\sum_{i,j}J_{ij}\mathbf S_i\cdot \mathbf S_j ,
\end{equation}
with spins on all the sites of a lattice but the $J_{ij}$  between neighbouring spins 
and chosen randomly from a distribution having weight of either sign.\footnote{In fact 
the EA Hamiltonian was first written explicitly by Sherrington and Southern (1975).}  
They further chose the distribution to be Gaussian of zero mean, thereby both 
eliminating the possibility of any conventional order (with spatially uniform or 
periodic magnetization) and also having a minimal one-parameter characterization\footnote{Actually, 
the Gaussian choice for the single parameter description was also useful for the further analytic 
methods employed. The alternative simple single-parameter symmetric distribution having two delta 
functions of equal weight at  $\pm J$ has often been employed in (later) computer simulations.}.  
This necessitated the introduction of a new form of order parameter to describe magnetic freezing 
without periodicity.  In fact EA gave two versions: one based explicitly on temporal freezing,
\begin{equation}
\label{eq3}
q=\lim_{t\to\infty,\tau\to\infty}q(t,t+\tau);\quad q(t,t+\tau)\equiv 
N^{-1}\sum_i\langle\mathbf S_i(t)\cdot\mathbf S_i(t+\tau)\rangle,
\end{equation}
where $\langle\cdot\rangle$ refers to a dynamical average, and the other based on ensemble-averaging,
\begin{equation}
\label{eq4}
q=N^{-1}\sum_i|\langle\mathbf S_i\rangle|^2,
\end{equation}
with $\langle\cdot\rangle$ now referring to an ensemble-average restricted to one 
symmetry-breaking macro-state. The phase transition is signalled by $q$ becoming 
non-zero. 

EA did not attempt a full solution but used several new variants of mean field theory, 
all requiring novel treatment beyond those conventional for a simple ferromagnet. The 
most sophisticated of them introduced and employed the so-called 
`replica trick' which replaces the average of the logarithm of the partition
function\footnote{The argument for studying the average of the logarithm of the partition function is that
the physical quantities it generates should be self-averaging, independent in the
thermodynamic limit of the specific instance of choice of the disorder.}, the physical
generating function\footnote{Certain physical observables can already be expressed as
derivatives of ln $Z$ with $Z$ defined as above with the bare $H$. Others, in principle,
require the addition to $H$ of terms involving appropriate conjugate fields so that
desired observables follow from derivatives of ln $Z$ with respect to these fields.} , by
the limiting behaviour of a partition function of an effective periodic system of higher
dimensional spins:
\begin{equation}
\label{eq5}
\overline{\hbox{ln}\,Z}=\lim_{n\to0}\partial /\partial n(\overline{Z^n})=\lim_{n\to0}\partial 
/\partial n(\overline{\prod_{\alpha=1,\dots,n}Z(\alpha)})=\lim_{n\to0}\partial
Z_{\text{eff}}(n)/\partial n ,
\end{equation}
where the overbar refers to the average over the distribution of the $J$, $Z$ is the usual 
partition function  $Z=\operatorname*{Tr}_{\{\sigma\}}\exp\{-\beta H\}$, $Z(\alpha)$ is 
the partition function for spins with dummy labels $\alpha$  and 
$Z_{\operatorname{eff}}(n)$ is the partition function of a periodic Hamiltonian 
$H_{\text{eff}}(\{\sigma_i^\alpha\})$ of effectively higher dimensional pseudo-spins with
extra replica labels $\alpha=1,\dots,n$   and higher order interactions now between spins
with different replica, as well as site, labels. Within this new description EA devised a
new mean field theory with a new order parameter measuring inter-replica overlap 
\begin{equation}
\label{eq6}
q^{\alpha\beta}=N^{-1}\sum_i\mathbf S_i^ \alpha\cdot
\mathbf S_i^\beta;\quad\alpha\not=\beta.
\end{equation}
To go further, however, they employed several assumptions and approximations whose validity was 
difficult to assess, although they do yield results with some qualitative similarity to several 
experimental features.

\subsubsection{Sherrington-Kirkpatrick}
In view of the many uncertainties of the EA analysis and the fact that the model was surely 
not soluble with current techniques, it seemed sensible to look for a model in which a 
mean-field theory might be exact. Since the conventional ferromagnet is soluble in the 
thermodynamic limit provided that all spins interact equally with one another and 
correspondingly the exchange interaction scales inversely with the number of spins, 
it seemed reasonable to look for an analogue in the spin glass problem. This led to 
the formulation of the Sherrington-Kirkpatrick (SK) model whose Hamiltonian is similar 
to that of EA but with interactions between all spins, chosen randomly and independently 
from a distribution whose mean and variance scale inversely with the number of spins. 
Simplifying to Ising spins and allowing for a ferromagnetic bias and an external field, 
the SK model is characterised by\footnote{We use notation $(ij)$ to denote a pair of unequal sites.}
\begin{equation}
\label{eq7}
H=-\sum_{(ij)}J_{ij}\sigma_i\sigma_j-h\sum_i\sigma_i;\quad\sigma=\pm1; \quad
J_{ij}\text{ i.i.d.; }\overline{J_{ij}^{}}=J_0 / N,\overline{J_{ij}^2}=J^2 / N.
\end{equation}
Despite its apparent simplicity, this model has turned out to expose many subtleties; 
for statistical physics, for mathematical physics and for probability theory; as well 
as having much wider application relevance.  Extensions to other related models with 
extensive and super-extensive constraints independently drawn from identical (intensive) 
distributions have led to further novelties and applications. In this introductory 
perspective I shall restrict discussion to outlines at the level of conceptual 
theoretical statistical physics, leaving mathematical rigour to other authors.

\subsubsection{Replica theory}
Within the replica theory of EA but applied to the SK model the averaged free energy 
can be expressed in a form 
\begin{equation}
\label{eq8}
\overline F=-T\overline{\operatorname{ln}Z}=-T\lim_{n\to0}\frac\partial{\partial n}
\operatorname*{Tr}_{\{\sigma_i^\alpha;\alpha=1,\dots,n\}}\exp \biggl\{ f\biggl(\sum_{i\alpha}\sigma_i^
\alpha,\sum_{i(\alpha\beta)}\sigma_i^\alpha\sigma_i^\beta\biggr)\biggr\},
\end{equation}
in which  $f$ involves the spin variables only in the form of sums over all sites and those sums 
only up to quadratic order. Hence, by introducing auxiliary (macroscopic) variables to linearize 
these sums, the trace over the spins may be taken to yield
\begin{equation}
\label{eq9}
\overline F=-T\overline{\operatorname{ln}Z}=-T\lim_{n\to0}\frac\partial{\partial n}\int 
\prod_{\alpha=1,\dots,n}dm^\alpha\prod_{(\alpha\beta)}dq^{\alpha\beta}\exp[-N\Phi(\{m^\alpha;q^{\alpha\beta}\})]
\end{equation}
with $\Phi$  intensive. Thus, provided the limit $n\to  0$ and the thermodynamic limit $N\to\infty$   
can be inverted, the method of steepest 
descents in principle yields a solution determined by an extremum of $\Phi$ . 
However, to take the limit $n\to  0$ an appropriate analytic form continuable to non-integer $n$ is 
needed and the correct way to achieve this is not obvious. EA and SK both used the natural 
`replica-symmetric' ansatze, 
\begin{equation}
\label{eq10}
m^\alpha=m,\text{ all }\alpha;\quad q^{\alpha\beta}=q,\text{ all }\alpha\not=\beta.\text{\footnotemark}
\end{equation}
\footnotetext{One also requires that the extremum of $\Phi$  be a minimum for the single-replica order 
parameter $m^\alpha$   but a maximum for the two-replica order parameter $q^{\alpha\beta}$.}
This ansatz also yields the identifications
\begin{equation}
\label{eq11}
m=\overline{\langle\sigma_i\rangle}\text{ and } q=\overline{|\langle \sigma_i\rangle|^2}.
\end{equation}
Already it gives many features qualitatively similar 
to ones found experimentally.
In fact, though, it does not in general give a stable solution (de Almeida \& Thouless 1978) and a 
much more subtle replica-symmetry-breaking ansatz for $q$ is needed to yield stability in all 
regions of control-parameter space. The Parisi ansatz (Parisi 1980) has satisfied this need and 
passed all subsequent stability tests.

Let me first describe Parisi's ansatz in terms of its original replica theory formulation 
and only turn later to its physical interpretation. $q^{\alpha\beta}$   
can be viewed as an $n\times n$ matrix with zeros on its 
diagonal elements\footnote{Some authors take $q^{\alpha\alpha}$   as unity (c.f. an extension of 
eqn. \eqref{eq6}) but here I am assuming the  $\alpha\alpha$  term is so taken explicitly.}.  The Parisi ansatz 
may be viewed as the result of a sequence of operations in which (i) $n (\equiv    m_0)$ is initially 
considered as an integer which is subdivided sequentially into an integral number of smaller intervals; first 
into $n/m_1$ blocks of size $m_1$, then each of the $m_1$ blocks into $m_1/ m_2$ blocks of size $m_2$ and so on 
sequentially, with all the $m_i$ integers and the ratios $m_i/m_{i+1}$ also integers, 
until $m_{k+1} =1$ (ii) $q^{\alpha\beta}$   is taken to have the value $q_i$ if 
$I(\alpha/m_i)=I(\beta/m_i)$, 
$I(\alpha/m_{i+1})\not =I(\beta/m_{i+1})$  where $I(x)$ is equal to the 
smallest integer greater than or equal to $x$, as illustrated below for 
the sequence $n=m_0=12; \quad m_1=4; \quad m_2=2; \quad m_3=1$

\[
q^{\alpha\beta}=
\left(
\raisebox{-82pt}{
\setlength{\unitlength}{14pt}
\begin{picture}(12.5,12)(0,0)
\put(0,0){\line(1,0){12}}
\put(0,0){\line(0,1){12}}
\put(12,12){\line(-1,0){12}}
\put(12,12){\line(0,-1){12}}

\put(0,4){\line(1,0){12}}
\put(4,0){\line(0,1){12}}
\put(0,8){\line(1,0){12}}
\put(8,0){\line(0,1){12}}

\put(0,10){\line(1,0){4}}
\put(2,8){\line(0,1){4}}
\put(4,6){\line(1,0){4}}
\put(6,4){\line(0,1){4}}
\put(8,2){\line(1,0){4}}
\put(10,0){\line(0,1){4}}

\put(0,11){\line(1,0){2}}
\put(1,10){\line(0,1){2}}
\put(2,9){\line(1,0){2}}
\put(3,8){\line(0,1){2}}
\put(4,7){\line(1,0){2}}
\put(5,6){\line(0,1){2}}
\put(6,5){\line(1,0){2}}
\put(7,4){\line(0,1){2}}
\put(8,3){\line(1,0){2}}
\put(9,2){\line(0,1){2}}
\put(10,1){\line(1,0){2}}
\put(11,0){\line(0,1){2}}

\put(2,2){\raisebox{-2pt}{\makebox[0pt][c]{$q_0$}}}
\put(6,2){\raisebox{-2pt}{\makebox[0pt][c]{$q_0$}}}
\put(2,6){\raisebox{-2pt}{\makebox[0pt][c]{$q_0$}}}
\put(10,10){\raisebox{-2pt}{\makebox[0pt][c]{$q_0$}}}
\put(6,10){\raisebox{-2pt}{\makebox[0pt][c]{$q_0$}}}
\put(10,6){\raisebox{-2pt}{\makebox[0pt][c]{$q_0$}}}

\put(3,11){\raisebox{-2pt}{\makebox[0pt][c]{$q_1$}}}
\put(7,7){\raisebox{-2pt}{\makebox[0pt][c]{$q_1$}}}
\put(11,3){\raisebox{-2pt}{\makebox[0pt][c]{$q_1$}}}
\put(1,9){\raisebox{-2pt}{\makebox[0pt][c]{$q_1$}}}
\put(5,5){\raisebox{-2pt}{\makebox[0pt][c]{$q_1$}}}
\put(9,1){\raisebox{-2pt}{\makebox[0pt][c]{$q_1$}}}

\put(1.5,11.5){\raisebox{-1pt}{%
               \makebox[0pt][c]{$\scriptstyle{q_2}$}}}
\put(3.5,9.5){\raisebox{-1pt}{%
               \makebox[0pt][c]{$\scriptstyle{q_2}$}}}
\put(5.5,7.5){\raisebox{-1pt}{%
               \makebox[0pt][c]{$\scriptstyle{q_2}$}}}
\put(7.5,5.5){\raisebox{-1pt}{%
               \makebox[0pt][c]{$\scriptstyle{q_2}$}}}
\put(9.5,3.5){\raisebox{-1pt}{%
               \makebox[0pt][c]{$\scriptstyle{q_2}$}}}
\put(11.5,1.5){\raisebox{-1pt}{%
               \makebox[0pt][c]{$\scriptstyle{q_2}$}}}
\put(0.5,10.5){\raisebox{-1pt}{%
               \makebox[0pt][c]{$\scriptstyle{q_2}$}}}
\put(2.5,8.5){\raisebox{-1pt}{%
               \makebox[0pt][c]{$\scriptstyle{q_2}$}}}
\put(4.5,6.5){\raisebox{-1pt}{%
               \makebox[0pt][c]{$\scriptstyle{q_2}$}}}
\put(6.5,4.5){\raisebox{-1pt}{%
               \makebox[0pt][c]{$\scriptstyle{q_2}$}}}
\put(8.5,2.5){\raisebox{-1pt}{%
               \makebox[0pt][c]{$\scriptstyle{q_2}$}}}
\put(10.5,0.5){\raisebox{-1pt}{%
               \makebox[0pt][c]{$\scriptstyle{q_2}$}}}
\end{picture}}
\right)
\]
(iii) in the limit $n\to0$  the $m$ are continued to real values with  
$0\le m_1\le m_2\le\dots\le m_k\le m_{k+1}\equiv1$ and $q$ is replaced by a function $q(x)$ given by 
$q(x)=q_i$; $m_i<x<m_{i+1}$ $(i=1,\dots,k)$  with $x$ in [0,1], (iv) the limit $k\to\infty$    is taken. 
Insertion into eqn.\ \eqref{eq9} yields a functional integral which in the limit $N\to\infty$ is extremally 
dominated and yields self-consistency equations for the dominating $q(x)$; hereafter $q(x)$ is taken to refer 
to this extremal function, which is the the mean-field order function for the problem. For different regions 
of the $(J,J_0,h,T)$  parameter space the stable solutions are of one of two forms:
\begin{enumerate}
\item[(i)] $q(x)=q=$ constant;  replica-symmetric (RS)
\item[(ii)] $q(x)=q_0$  for $0\le x\le x_1$, monotonically increasing smoothly 
between $x_1$ and $x_2$, and $q(x)=q_1$ for $x_2\le x\le 1$; full replica symmetry breaking (FRSB)
\footnote{For the intermediate approximations mentioned above one would have a $k$-step replica symmetry-breaking 
with $k+1$ sections of constant $q(x)$ separated by $k$ discontinuities, but it is believed that the only stable 
situations for the SK model are $k=0$ (RS) and $k=\infty$ (FRSB). There are stable 1 step RSB solutions for several 
other problems (see section 5).}.
\end{enumerate} 
They are separated by a plane in $(J,J_0,h,T)$ which marks a continuous transition, with RSB on the higher-$J$ side.

Replica symmetry breaking signals the existence of many non-equivalent macrostates. 
$q(x)$ provides a measure of the extent of similarity between these states. It follows 
from consideration of the concept of overlaps (Parisi 1983). The overlap between two macrostates $S$, 
$S'$ is defined by
$q^{SS'}=N^{-1}\sum_i\langle\sigma_i\rangle^S\langle\sigma_i\rangle^{S'}$,
 where  $\langle\cdot\rangle^S$ refers to a thermodynamic average over macrostate $S$,  and the
distribution of overlaps is given by $P(q)=\sum_{S,S'}W_SW_{S'}\delta(q-q^{SS'})$ where $W_S$
is the probabilistic weight of state 
$S$, given in equilibrium by $W_S=\exp(-\beta F_S)/\sum_{S'}\exp(-\beta F_{S'})$ where 
$F_S$ is the free energy of macrostate $S$. The relation to $q(x)$ is 
$\overline{P(q)}=\int dx\delta(q-q(x))=dx/dq$. Consequently it follows that an RS system has a 
single macrostate (aside from trivial global inversion or rotation), whereas FRSB implies a 
hierarchy of non-equivalent relevant macrostates at the temperature of interest\footnote{There 
are even more macrostates of relevance at different temperatures.}. This in turn 
implies that the macroscopic dynamics will be slow and glassy and that practical 
equilibration will be very difficult to achieve. Already, however, the existence of RSB predicts 
different kinds of response functions; 
for the susceptibility one may experience either single-macrostate response
$\chi_{SS}\equiv\chi_{EA}=T^{-1}(1-q(1))$  or the full Gibbs average $\chi_G=T^{-1}(1-\int
dxq(x))$. These in turn can be identified with the experimental zero-field-cooled and
field-cooled susceptibilities and used to explain their 
difference in the spin glass phase (see e.g. Nagata et al. 1979); this non-ergodicity 
was already observed before EA
in the difference between thermoremanent and isothermal remanent magnetisations 
(e.g. Tholence and Tournier 1974).  The
Parisi replica analysis also demonstrates a number of other interesting properties 
(M\'ezard et al.\ 1984), such as ultrametricity (M\'ezard and Virasoro 1985)\footnote{This 
corresponds to the hierarchical clustering of overlaps as illustrated
by the branching cartoon 
\[
\raisebox{-27pt}{
\setlength{\unitlength}{8pt}
\begin{picture}(11,7)(0,0)

\qbezier(0,1)  (0,2)  (0.4,3)
\qbezier(0.1,1)(0.1,2)(0.5,3)
\qbezier(0.9,1)(0.9,2)(0.5,3)
\qbezier(1,1)  (1,2)  (0.6,3)

\qbezier(2,1)  (2,2)  (2.4,3)
\qbezier(2.1,1)(2.1,2)(2.5,3)
\qbezier(2.9,1)(2.9,2)(2.5,3)
\qbezier(3,1)  (3,2)  (2.6,3)

\qbezier(4,1)  (4,2)  (4.4,3)
\qbezier(4.1,1)(4.1,2)(4.5,3)
\qbezier(4.9,1)(4.9,2)(4.5,3)
\qbezier(5,1)  (5,2)  (4.6,3)

\qbezier(6,1)  (6,2)  (6.4,3)
\qbezier(6.1,1)(6.1,2)(6.5,3)
\qbezier(6.9,1)(6.9,2)(6.5,3)
\qbezier(7,1)  (7,2)  (6.6,3)

\qbezier(8,1)  (8,2)  (8.4,3)
\qbezier(8.1,1)(8.1,2)(8.5,3)
\qbezier(8.9,1)(8.9,2)(8.5,3)
\qbezier(9,1)  (9,2)  (8.6,3)

\qbezier(10,1)  (10,2)  (10.4,3)
\qbezier(10.1,1)(10.1,2)(10.5,3)
\qbezier(10.9,1)(10.9,2)(10.5,3)
\qbezier(11,1)  (11,2)  (10.6,3)

\qbezier(0.4,3)(0.4,3.5)(1.3,4.5)
\qbezier(0.6,3)(0.6,3.5)(1.5,4.42)
\qbezier(2.4,3)(2.4,3.5)(1.5,4.42)
\qbezier(2.6,3)(2.6,3.5)(1.7,4.5)

\qbezier(4.4,3)(4.4,3.5)(5.3,4.5)
\qbezier(4.6,3)(4.6,3.5)(5.5,4.42)
\qbezier(6.4,3)(6.4,3.5)(5.5,4.42)
\qbezier(6.6,3)(6.6,3.5)(5.7,4.5)

\qbezier(8.4,3)(8.4,3.5)(9.3,4.5)
\qbezier(8.6,3)(8.6,3.5)(9.5,4.42)
\qbezier(10.4,3)(10.4,3.5)(9.5,4.42)
\qbezier(10.6,3)(10.6,3.5)(9.7,4.5)

\qbezier(1.3,4.5)(2,5.8)(4.9,6.5)
\qbezier(1.7,4.5)(2.4,5.7)(5.3,6.3)
\qbezier(5.3,4.5)(5.3,4.5)(5.3,6.3)
\qbezier(5.7,4.5)(5.7,4.5)(5.7,6.3)
\qbezier(9.3,4.5)(8.6,5.7)(5.7,6.3)
\qbezier(9.7,4.5)(9,5.8)(6.1,6.5)
\qbezier(4.9,6.5)(4.9,6.5)(4.9,7.)
\qbezier(6.1,6.5)(6.1,6.5)(6.1,7.)

\put(-0.5,0){\makebox[0pt][r]{$\scriptstyle{\alpha =}$}}

\put(0,0){\makebox[0pt][c]{$\scriptstyle{1}$}}
\put(1,0){\makebox[0pt][c]{$\scriptstyle{2}$}}
\put(2,0){\makebox[0pt][c]{$\scriptstyle{3}$}}
\put(3,0){\makebox[0pt][c]{$\scriptstyle{4}$}}
\put(4,0){\makebox[0pt][c]{$\scriptstyle{5}$}}
\put(5,0){\makebox[0pt][c]{$\scriptstyle{6}$}}
\put(6,0){\makebox[0pt][c]{$\scriptstyle{7}$}}
\put(7,0){\makebox[0pt][c]{$\scriptstyle{8}$}}
\put(8,0){\makebox[0pt][c]{$\scriptstyle{9}$}}
\put(9,0){\makebox[0pt][c]{$\scriptstyle{10}$}}
\put(10,0){\makebox[0pt][c]{$\scriptstyle{11}$}}
\put(11,0){\makebox[0pt][c]{$\scriptstyle{12}$}}

\end{picture}}
\qquad
\left\{
\begin{array}{l}
q^{(1\,5)}=q^{(1\,6)}=q^{(1\,7)}=\ldots = q_0 \,; \\
q^{(1\,3)}=q^{(1\,4)}=q^{(2\,3)}=\ldots = q_1 \,; \\
q^{(1\,2)}=q^{(3\,4)}=q^{(5\,6)}=\ldots = q_2 \,;
\end{array}
\right.
\]
}
and
non-self-averaging of certain non-trivial overlap measures. but these will not be dwelt upon
further here\footnote{For a recent review of the topic of overlaps and their
interpretation see (Parisi 2004).}.

\subsubsection{Short-range spin glasses}
`Real' experimental spin glasses have short-range or spatially decaying exchange
interactions, whereas the replica theory above is exact only for infinite-range problems.
Many of the predictions of mean field theory are mimicked qualitatively in the
experiments; some are thought to be real, but others are still subjects of controversy in
true Gibbs equilibrium although often apparent as non-equilibrium experimental features.
The Edwards-Anderson model with nearest neighbour interactions is considered
representative of such real spin glasses but remains without full exact solution.

\subsubsection{Spin glasses on dilute random networks}
A class of model spin glasses with finite inter-spin connectivities, as is the case for
EA, but range-free and offering the possibility of exact solution, was introduced by
Viana \& Bray (1985) and characterised by an analogue of SK with
\begin{equation}
\label{eq12}
H=-\sum_{(ij)}c_ic_j J_{ij}\sigma_i\sigma_j;\text{ random quenched $c$ 
and $J_{ij}$; }c_i=0,1;\ \overline{J_{ij}}=J_0,\ \overline{J_{ij}^2}=J^2,
\end{equation}
where the annealled spins $\sigma$  are located on the quenched vertices of a finite-connectivity 
Erd\"os-Renyi\footnote{In an  Erd\"os-Renyi graph of degree $p$ any vertex is 
connected to any other with a probability $p/N$.}  graph, with $c_i=1$  denoting a vertex, 
but without the need for inverse $N$-scaling of the exchange distribution.
 \footnote{A simple extension 
utilises as underlying network a random graph with fixed degree at each 
vertex 
(Banavar et. al. \ 1987).}. This problem, which is often considered a `half-way
 house' between
SK and EA, requires more order parameters 
$m^\alpha$, $q^{\alpha\beta}$, $q^{\alpha\beta\gamma}$, $q^{\alpha\beta\gamma\delta},\dots$  
and, although soluble
in RS approximation via a mapping $q^{\alpha\beta\dots r}=\int P(h)\{\operatorname{tanh}(\beta h)\}^r$, 
also poses greater challenges than SK for FRSB (see e.g. Wong and Sherrington 1988).

\subsubsection{Itinerant spin glasses}
Thus far we have discussed only systems with magnetic moments even in the absence of interaction. 
However, it is well known that ferromagnetism in periodic systems can occur not only through the 
orientation of effectively pre-existing localized moments, as typified by Curie-Weiss mean
field theory and found in insulating magnets and in some rare earth metals, but also through
the spontaneous cooperative ordering of metallic itinerant electrons, as in
Stoner-Wohlfarth ferromagnetism in transition metals. Similarly, one can readily
envisage itinerant spin glass behaviour (Sherrington and Mihill 1974) and indeed it is
found in alloys such as \textbf{Rh}Co (Coles et.al.\ 1974). A simple model is given by the
Hamiltonian 
\begin{equation}
\label{eq13}
H=\sum_{ij\sigma}t_{ij}a_{i\sigma}^+a_{j\sigma}+\sum_i V_i a_i^+a_i+\sum_i U_in_{i\uparrow}n_{i\downarrow},
\end{equation}
where the $a_i$, $a_i^+$ are Wannier electron creation and annihilation operators, 
$n_{i\sigma}\equiv a_{i\sigma}^+a_{i\sigma}$ are number operators, and the parameters 
$t_{ij}$, $V_i$, $U_i$  depend upon the types of atom at sites $i,j$. The simplest instance takes 
randomly quenched alloys with two atomic types $(A,B)$ but with the $t_{ij}$ independent of the atom 
types and considers only magnetic fluctuations
\begin{equation}
\label{eq14}
H=\sum_{ij\sigma}t_{ij}a_{i\sigma}^+a_{j\sigma}-\frac12\sum_i U_i(n_{i\uparrow}-n_{i\downarrow})^2,
\end{equation}
in which the $U_i$ take one value $U_A$ at the sites associated with atom type A and take the 
another value $U_B$ at the sites associated with atom type B. Of particular interest is the itinerant case in 
which (i) A is not spontaneously magnetically ordered, i.e. $(1-U_A\chi^0(q))>0)$ where $\chi^0(q)$ is the
wave-vector dependent susceptibility associated with the bare band structure, (ii) the pure B system is 
spontaneously itinerantly ferromagnetic (so $(1-U_B\chi^0(0))<0)$, but also (iii) there is no magnetic 
moment associated with an isolated B atom in an A matrix, even in the mean field sense of Anderson (1961). 
Analogy with the phenomenon of Anderson localization (Anderson 1958) leads to the expectation of statistical 
fluctuation nucleation of cluster moments within the conceptual framework of Anderson local moment 
formation, while further cluster interaction can stabilise cluster glass behaviour beyond a 
critical B concentration  (Sherrington and Mihill 1974), as well as ferromagnetism at a higher concentration. 
However, in fact isolated paramagnetic cluster moments are not necessary precursors for spontaneous 
spin glass order (as emphasised by Hertz 1979), as neither there are local moments in pure itinerant 
ferromagnets nor well-defined bosons in BCS superconductivity above their respective onset temperatures.   
 
A classical mean field theory follows from (14) by formulating the partition function as a functional
integral over a Grassmann representation of the electron field (Sherrington 1971), linearizing
the interaction term involving the $U$ over auxiliary Hubbard-Stratonovich fields, integrating out the 
electron fields and taking the static approximation. This yields an effective classical field theory with
\begin{equation}
\label{eq15}
Z=\int D\bf{m} \exp(-\beta F({\bf{m}})),
\end{equation}  
where
\begin{equation}
\label{eq16}
F({\bf{m}})=\sum_{i}{U_{i}m_{i}^2} - \sum_{ij}{U_{i}U_{j}m_{i}m_{j}\chi_{ij}^0} 
-\sum_{ijkl}{U_{i}U_{j}U_{k}U_{l}m_{i}m_{j}m_{k}m_{k}\Lambda_{ijkl}^0} + ...,
\end{equation}
where $m_{i}=<n_{i\uparrow}-n_{i\downarrow}>$ and $\chi_{ij}^0$ , $\Lambda_{ijkl}^0$, ... are two-, four- and 
..-point correlation functions of the bare band structure (in real space). Taking the extremum
yields a set of self-consistent mean field equations which are the analogue of the 
Thouless-Anderson-Palmer (1977) (TAP) equations for the SK equation. The analogy with Anderson localization
follows from writing these equations as
\begin{equation}
\label{eq17}
U_{i}^{-1}M_i - \sum_{j}{\chi_{ij}^{0}M_j} -\sum_{jkl}{\Lambda_{ijkl}^{0}M_{j}M_{k}M_{l}} +..=0; M_i=U_{i}m_i
\end{equation}
and comparing with the Anderson wave-function localization equation
\begin{equation}
\label{eq18}
\epsilon_{i}\phi_{i} + \sum_{j}{t_{ij}{\phi_j}} - E\phi_{i} = 0,
\end{equation}
with disorder in the $\epsilon_i (= U{_i}^{-1})$; naively, local moment clusters of (17) are 
related to negative energy states 
of (18) and long range magnetic order is related to the mobility edge. But in fact there are more subtle 
effects, both 
bootstrap effects as mentioned earlier (contained in the non-linear terms of (17)) and effects
differentiating spin glass
and ferromagnetic cooperative order.

A simple conceptual model of itinerant spin glass ordering, further simplified in the EA spirit,
is given by an effective field theory with 
\begin{equation}
\label{eq19}
Z=\int{\prod_{i}{d\phi_{i}}\exp(-F({\phi_{i}}))};F({\phi})=r\sum_{i}{\phi_{i}}^2 + u\sum_{i}{\phi_i}^4 
-\sum_{(ij)}J_{ij}\phi_{i}\phi_{j};u>0,
\end{equation}
with the $J_{ij}$ random as in EA or SK; this model encompasses local moment spin glasses 
for $r<0$ and itinerant spin glasses for $r>0$.

\subsubsection{Other induced moment models}
There are other classical models allowing the bootstrapping of magnetic order. One such 
is the spin glass analogue of the induction of ferromagnetism due to exchange interaction lifting 
of singlet ground state preference of isolated atoms. A simple example is the spin-1 Ising model
\begin{equation}
\label{eq20}
H=-D\sum_{i}S_i^2-\frac12\sum_{ij}J_{ij}S_iS_j;\quad S_i=0,\pm1.
\end{equation}
If $D>0$ then the system behaves analogously to the usual spin $1/2$ Ising model, but if 
$D<0$
then in the absence of $J$ the ground state preference is for non-magnetic $S_i=0$.
However even if $D<0$  a sufficient exchange can self-consistently lift the preference to
the magnetically ordered state via a first-order transition. If the $J_{ij}$ are quenched
random as in the SK model, this system is known as the Ghatak-Sherrington (GS) model
(1977) and has induced spin glass behaviour; it has been analysed 
extensively in FRSB 
by Crisanti and Leuzzi (2002). The Fermionic Ising Spin Glass (FISG) model (Rosenow and 
Oppermann 1996) is closely related (P\'erez Castillo and Sherrington 2005).

\subsubsection{Vector spin glasses}
Magnetic alloy spin glasses are not restricted to Ising systems. Indeed Heisenberg 
magnets are more common experimentally. It is straightforward to extend the exactly
soluble models to encompass vector spins (see e.g. Sherrington 1983). In the absence 
of a magnetic field or a
ferromagnetic component there is little change of note beyond the extension of random
spin glass freezing to the full spin dimensionality\footnote{But it might be noted that 
the choice of Ising spins by SK led to the `smoking gun' of negative entropy at T=0 and
the realisation that there was a 
subtlety, which eventually led to Parisi's ansatz and beyond; negative entropy at T=0 was
a known pathology of continuous classical spins but should not occur for discrete spins 
such as Ising.}. Within the infinite-range/ mean
field model,  an axial symmetry-breaking due to an applied field or to ferromagnetism still permits a
spin glass freezing in the orthogonal directions (Gabay and Toulouse 1981) with strong
onset of transverse non-ergodicity and induced weaker longitudinal non-ergodicity,
crossing over to strong RSB in all directions at a lower temperature (Cragg et a.\ 1982,
Elderfield and Sherrington 1982, 1984). Anisotropy effects can also be included (Cragg and
Sherrington 1982b).

\section{Discontinuous transitions}
For the case of conventional 2-spin interactions, as employed in both the SK and EA
models and believed to be appropriate for conventional experimental magnetic alloy spin
glasses, mean field theory yields full replica symmetry breaking once the spin glass
state occurs\footnote{Although it remains controversial as to whether any RSB holds in
short-range systems}. However, in extensions which lack reflection and definiteness
symmetries, such as $p$-spin models for $p>2$ (Crisanti and Sommers 1992) or Potts or
quadrupolar spin glasses beyond critical Potts or vector dimensions  (Gross et al 1985, 
Goldbart and Sherrington 1985) one finds that 
the spin glass transition is
discontinuous to one step of replica symmetry breaking with finite overlap magnitude
(D1RSB)\footnote{In a Potts or quadrupolar model 
for a range of intermediate Potts or vector dimensions the transition to 1RSB 
is continuous (Elderfield and
Sherrington 1983, Goldbart and Sherrington 1985, Sherrington 1986); a similar 
transition to C1RSB occurs in 
a $p>2$-spin model in a sufficient applied field (Crisanti and Sommers 1992). Except
for spherical spins, there is also a lower temperature transition from 
1RSB to FRSB (Gardner 1985, Gillin et al (2001).}. This behaviour is
thought to be characteristic also of (even short-range interaction) structural glasses,
in which crystallization is dynamically avoided in favour of self-consistent glassiness.

\section{Beyond magnetic alloys}

\subsection{Complex many body problems}
The formalism and concepts developed for model magnetic alloys have found significant
application more generally; in particular for a large class of problems that can be
characterised by control functions of the form
\begin{equation}
\label{eq21}
H=H(\{J_{ij\dots k}\},\{S_{ij\dots l}\},\{X\}),
\end{equation}
where the $i$, $j$ are microscopic identification labels; the
$\{J_{ij\dots k}\}$ symbolise a set of quenched parameters depending on one or more of the
identification labels and in general different for different labels; the $\{S_{ij\dots
l}\}$ symbolise the (annealled) microscopic variables again depending on one or more
identification variables; and the  $\{X\}$ are macroscopic intensive control variables. 
The specific identifications of the
$\{J,S,X\}$ can however be quite different, as also the manner of operation of the control
function. In the spirit of statistical physics and probability theory one often concerns oneself
with problems in which the $\{J,X\}$ are drawn from intensive distributions
independent of the specific labels.

\subsubsection{Examples}
We have already seen one example in the case of a magnet with the $i$ labelling the spins,
the $J$ exchange interactions, the $S$ spin orientations, $X$ the temperature and $H$ the
Hamiltonian determining the distribution of the $S$ through the Boltzmann measure. Other
examples include: 

\begin{enumerate}
\item[(i)] \textit{The Hopfield neural network:}
Here the $i$ label neurons,  $\{S_i\}$ indicate the states of the neurons as firing or not
firing, $\{J_{ij}\}$  label synaptic efficacies given in terms of (randomly chosen quenched) 
stored patterns
$\{\xi_i^\mu\}$; $\mu=1,\dots,p=\alpha N$  by
$J_{ij}=N^{-1}\sum_{\mu}\xi_i^\mu\xi_j^\mu$, $X\equiv T\equiv\beta^{-1}$ is a measure
of the rounding of the sigmoidal response of a neuron to the sum of its incoming signals,
$H=-\frac12\sum_{ij}J_{ij}S_i S_j$ and $P(\{S\})\sim\exp(-\beta H)$  characterises the
stationary macro firing states. From the neural retrieval perspective, 
however, interest is not in the full
Gibbs average but rather in the individual retrieval macrostates with macroscopic
overlaps $m^\mu=N^{-1}\sum_i\xi_i^\mu\langle S_i\rangle$ with the patterns coded in the
$\{J\}$; retrieval corresponds to a finite overlap with a single pattern and is an
analogue of ferromagnetism in the examples of section 2. Spin glass states do occur due
to pattern interference but are not the desired states in neural operation and their
dominance indicates breakdown of retrievable memory\footnote{For further
discussion see for example Sherrington
(1992) or Nishimori (2001).}.
\item[(ii)] \textit{Hard optimization:} 
Here the objective is to minimise a cost function $H$ as a function of variables $\{S\}$
with constraints $\{J\}$. An example is the problem of partitioning the vertices $i$ of a
random graph into two groups of equal size but with the minimum number of edges of the
graph between the two groups. This can be formulated as finding the ground state of a
Viana-Bray-like spin glass. Consequently it can be studied by an analogue of the procedure of
studying the thermodynamics of the VB spin glass. If the interest is in finding the
average minimum spanning cut then replica procedure may be employed, inventing an
artificial annealing temperature $T$ and taking it to zero at the end of the calculation.
Of course the actual calculation involves all the subtleties of replica symmetry breaking
and computer simulation involves all the corresponding issues of slow glassy 
dynamics\footnote{Simulation also exhibits the spin-glass features of ultrametricity 
and non-self-averaging (Banavar et al 1987).}.
Another optimization problem in artificial neural network theory is to determine the
maximum number of patterns which can be stored and retrieved with a specified maximum
error; in this case the variables are the synaptic efficacies and the quenched parameters
are the stored patterns.
More recently many other optimization problems have been studied by techniques derived 
from spin glass studies.
\item[(iii)] \textit{Error-correcting codes:}
One procedure for coding and retrieval is to code the information 
to be transmitted in the form of exchange interactions whose insertion into an effective
magnetic Hamiltonian yields a ground state which identifies the desired message. In
practice, however, transmission lines add noise and retrieval is required to best
eliminate the effects of the noise. This yields yet another optimization problem, with
best retrieval resulting from the introduction of an effective retrieval temperature-noise 
matching 
that on the line\footnote{Again see Nishimori (2001) for further details.}.
Indeed there are several other problems in which the optimal character of noise matching
can be demonstrated.
\end{enumerate}

\section{Dynamics}
Thus far discussion has been about equilibrium or quasi-equilibrium. 
However, often one wishes to consider dynamics, including away from 
equilibrium;\footnote{Note that whereas in real 
physical situations the true microscopic dynamics is determined
by nature, in computer simulations the dynamics is chosen by the simulator and there 
exists
the opportunity to optimise that choice. Similarly, the control fields $X$ are choosable.},
indeed if detailed balance is not present one cannot use usual Boltzmann equilibrium
theory. As before, we are normally interested in systems characterised by simple
distributions. Again one can utilise the general picture of a controlling function as in
(21) but now operating in an appropriate microscopic dynamics (and without necessarily
symmetries such as $J_{ij}=J_{ji}$). The analogue of the use of the partition function
for thermodynamics is to use a dynamical generating functional (de Dominicis 1978) which can be expressed
symbolically either, for random sequential updates, as 
\begin{equation}
\label{eq22}
Z(\Lambda)=\int D\mathbf S(t)\Pi\delta\text{ (eqn of motion) }\exp\biggl(\int dt\Lambda(t)
\cdot\mathbf S(t)\biggr),
\end{equation}
where the
integral is over all variable paths in the full space-time, the $\Pi\delta$ term indicates that
the microscopic equations of motion are always satisfied and the 
$\Lambda\cdot\mathbf S$ term is a generating term, or, for parallel updates, as
\begin{equation}
\label{eq23}
Z(\Lambda)=\int \Pi d\mathbf S(t)\prod_t W(\mathbf S(t+1)|\mathbf S(t))
P_0(\mathbf S)\exp\biggl(\sum_{t}\Lambda(t)\cdot\mathbf S(t)\biggr).
\end{equation}
where $P_t(\mathbf S)$ indicates the ensemble distribution of $\mathbf S$ at time $t$ and 
$W(\mathbf S(t+1)|\mathbf S(t))$ indicates the probability of updating from 
$\mathbf S(t)$ to
$\mathbf S(t+1)$. With suitable Jacobian normalization (not shown explicitly) 
$Z(\Lambda=0)=1$  and one can average over the
quenched disorder without need for replicas; instead of interactions between replicas one
gets effective interactions between different epochs. In the case of range-free problems
one can again eliminate microscopic variables in place of macroscopic ones by the
artifice of introducing new variables via relations such as
\begin{equation}
\label{eq24}
\begin{split}
1&=\int dC(t,t')\delta(C(t,t')-N^{-1}\sum_i S_i(t)S_i(t'))\\
&=\int d\hat C(t,t')dC(t,t')\exp\biggl\{i\hat C(t,t')(C(t,t')-N^{-1}\sum_i
S_i(t)S_i(t'))\biggr\}
\end{split}
\end{equation}
and similarly for
response functions (involving also operators corresponding to $\partial/\partial
S_i(t)$). One can then integrate out the microscopic variables to leave purely
macroscopic measures; in the simplest cases of the form
\begin{equation}
\label{eq25}
Z_{\text{eff}}\sim\int D\tilde{\mathbf C}(t,t',t'',\dots)
\exp(N\Phi(\{\tilde{\mathbf C}\})),
\end{equation}
where $\tilde{\mathbf C}$ is used to denote the generic set and
the temporal dependence is two-time for full connectivity of the SK type but includes all
numbers of different times for VB finite connectivity. Steepest descents then yields
self-consistent coupled equations for the macroscopic correlation and response functions,
although of course boundary conditions need care. This is the dynamical analogue of
replica thermodynamics. In general, however, it is more difficult than replica theory
and fewer cases have been solved fully. Also, in some cases the convenience of a final
expression purely in terms of coupled correlation and response functions is not 
available, although
alternative descriptions in terms of ensembles of effective single agents can often be obtained.

An alternative procedure invoking an infinite multiplicity of single-time order 
parameters has also been considered but will not be pursued here (see e.g. Coolen et al. 1996)

\subsection{Examples}

\subsubsection{$p$-spin spherical spin glass}
One example of the above procedure has been the analysis of the (infinite-range) $p(>2)$-spin
spherical spin glass, of Hamiltonian
\begin{equation}
\label{eq26}
H=-\sum_{i_1<i_2<\dots<i_p}J_{i_1i_2\dots i_p}S_{i_1}S_{i_2}\dots
S_{i_p};\quad\sum_i S_i^2=N;\quad\ 
\overline{J_{i_{1}i_{2}..i_{p}}^2}=J^2p!/2N^{p-1} 
\end{equation}
and obeying Langevin dynamics, for which closed equations in terms of correlation and response functions have been
obtained (Cugliandolo \& Kurchan 1993). In general these equations are not restricted to
stationarity. Analysis has indicated that above a critical temperature, known as the
dynamical transition temperature, stationary solutions do exist, with $\tilde C(t,t')\equiv
\tilde C(t-t')$ and satisfying the normal fluctuation-dissipation theorem and mode-coupling
theory, but below this temperature equilibration does not occur, the normal
fluctuation-dissipation theorem $-dR/dC=\beta$ (where $R$ is the integrated response, $C$
is the correlation function and $\beta$ is the inverse temperature) is replaced by a
modified relation $-dR/dC=\beta X(C)$  where $X(C)=x(C)$ with $x(q)$ the inverse of the
Parisi function $q(x)$, the $R$ and $C$ are now two-time (and non-stationary) and the
$C$-dependence of $X(C)$ is instantaneous-parenthetic\footnote{See for example Parisi
(2004)} \footnote{In this case the onset of RSB is discontunuous and the transition 
temperature is given differently by simple extremization of replica theory and dynamically.
The correct comparison with dynamics within replica theory is determined using marginal
stability.}. 

These and related dynamical studies vindicate and quantify the concepts of glassy 
dynamics deduced from the thermodynamic existence of many non-equivalent metastable
macrostates and the barriers between them.   

\subsubsection{Dynamical SK-model}
In the $p(>2)$-spin spherical spin glass model there is only one step of 
RSB in the replica equilibrium theory and similarly only 2 straight slope regions for
$X(C)$. The $p=2$ SK Ising system is more complicated with more structure, corresponding
to the hierarchy of FRSB, and dynamical analogues of ultrametricity (Cugliandolo and
Kurchan 1994). Other models can show regions of 1-RSB and of FRSB thermodynamics 
\footnote{For example the $p>2$-spin Ising model has 1-RSB thermodynamic behaviour above 
a critical temperature 
but then FRSB behaviour below; see {\it e.g.} Gardner (1985), Gillin et.al. (2001).} while 
it seems likely that dynamical vestiges of FRSB may occur in many systems, 
even with 1RSB thermodynamics. 

\subsubsection{Minority game}
A rather different example is found in the so-called Minority Game in econophysics 
(see {\it e.g.} \ Challet et al.\ 2004, Coolen 2004), which mimics a system of speculative agents
in a model market trying to gain by minority action. In the batch version of this game
the system obeys microdynamics
\begin{equation}
\label{eq27}
p_i(t+1)=p_i(t)-h_i-\sum_{j}J_{ij}\operatorname{sgn}p_j(t);\quad
h_i=N^{-1}\sum_j\vec\xi_i\cdot\vec\omega_j;\quad J_{ij}=N^{-1}\vec\xi_i\cdot\vec\xi_j ,
\end{equation}
with the $i$ labelling agents, the $p$ unbounded variables corresponding to strategy
point-weightings, and the $\vec\xi$ and $\vec\omega$  quenched random vectors in a
$D$-dimensional strategy space. This system is soluble along the lines outlined above,
utilizing large-$N$ steepest descent domination, in terms of an ensemble of independent
agents obeying non-Markovian stochastic dynamics with ensemble-self-consistently
determined coloured noise. On the macroscopic level it exhibits an ergodic-nonergodic
transition at a critical value $d_c$ of $d=D/N$, asymptotically independent of
preparation for $d>d_c$ but preparation-dependent for $d<d_c$.

\section{Conclusion}
The spin glass problem
has yielded many new concepts and techniques in both theoretical and experimental
physics. These concepts and techniques have in turn inspired new insights and practical
opportunities in the wider field of complex many-body problems, ranging through physics,
computer science, biology and economics, with pastures still open in these and the social
sciences. Most of this work has been on simple models with a single level of microscopic
timescale (but many resulting macro timescales) but some work has started and much
remains to do when different parameters are allowed different microdynamic time-scales
(as for example in neural networks where both neurons and synapses evolve, the former
much faster than the latter, or in biological evolution where the timescales of organism
operation and species evolution and mutation are very different). Although physical
systems normally obey detailed balance, others need not (e.g.\ biological or economic or
social systems). Most of the theoretical work has been performed at a level of uncertain
if physically reasonable approximation or assumption. Greater mathematical physics rigour
is now needed and will be the topic of other authors in this volume. The spin glass
models have introduced also new concepts in probability theory that are stimulating new
mathematics. Spin glass dynamics poses challenges yet to be investigated with
mathematical rigour. Much has been achieved but much remains to do. 

\section{Acknowledgements}
My career in spin glasses has involved and benefited from many valuable collaborations 
and discussions with too many people to list, but they know who they are and I thank them all. 
I have also benefited from financial support from UK Engineering and Physical Sciences Research 
Council (and its predecessors), the European Science Foundation, the European Community and 
the Royal Society, as well as from Imperial College, the University of Oxford and other 
universities and institutions. 

I would like to thank explicitly Dr Isaac P\'erez Castillo, 
for a careful reading and correction of drafts of this paper and Dr Andrea Sportiello for 
providing the drawings.

\end{document}